\documentstyle[12pt]{JHEP3}
\newcommand{\be}[1]{ \begin{equation}\label{#1} }
\newcommand{\ee}{\end{equation}}
\newcommand{\bea}[1]{\begin{eqnarray}\label{#1} }
\newcommand{\eea}{\end{eqnarray}}
\newcommand{\eq}[1]{(\ref{#1})}
\def\ZZZ{{\hskip-3pt\hbox{ Z\kern-1.6mm Z}}}
\def\zzz{{\hskip-3pt\hbox{ z\kern-1mm z}}}

\def\one{{\hbox{ 1\kern-.8mm l}}}
\def\zero{{\hbox{ 0\kern-1.5mm 0}}}

\title{Correlators in the Simplest Gauge-String Duality}

\author{
Rajesh Gopakumar, Roji Pius $^a$ \\ 
$^a$Harish-Chandra Research Institute, \\
$\;$Chhatnag Road,\\
$\;$Jhusi, India 211019\\
$\;$\email{gopakumr at hri.res.in, rojipius at hri.res.in}}

\abstract{In this note we compare planar correlators such as $\langle \prod_i^n {\rm Tr} M^{2k_i} \rangle_{conn}$ in the Gaussian matrix model with corresponding genus zero correlators of the A-model topological string theory on ${\mathbb P}^1$. We find a simple relation between them which provides additional evidence for the duality between the two theories, as proposed in \cite{Gopakumar:2011ev}.}

\preprint{HRI/ST/1213}

\begin{document}

\section{Introduction}

A simple example of how the Feynman diagrams for an $n$-point gauge correlator glue up into 
an $n$-point string scattering amplitude in a dual space-time can potentially provide a lot of insight into how gauge-string duality works. 

A candidate proposal for the ``simplest gauge-string duality" was put forward in \cite{Gopakumar:2011ev}. It relates the Gaussian one matrix integral in a large $N$ 'tHooft limit to the A-model topological string theory on ${\mathbb P}^1$ \cite{Witten:1989ig, Dijkgraaf:1990nc}. Gauge invariant correlators of the single trace operators ${\rm Tr} M^{p}$ can then be expected to be related to physical vertex operator scattering amplitudes in the dual topological string theory. 

Two pieces of evidence in favor of this (for planar connected correlators of  ${\rm Tr} M^{2p}$) were discussed in \cite{Gopakumar:2011ev}. The first was a nontrivial agreement of the degree of the covering map (from the worldsheet ${\mathbb P}^1$ to the target ${\mathbb P}^1$) which contributes to a given correlator. The second was a matching of the two point function
$\langle {\rm Tr} M^{2k_1} {\rm Tr} M^{2k_2} \rangle_{conn}$ (for arbitrary $k_1, k_2$). While this was encouraging, one needs stronger checks. Fortunately, it is possible to carry these out explicitly for a large class of correlators and this is the aim of this note.  We find, from the explicit computations on both sides of the putative duality, that there is a simple relation between the two sides which is a natural realization of gauge-string duality in this context. 

Before outlining this relation, we briefly recap the thread of logic followed in \cite{Gopakumar:2011ev}. The starting point was the observation by de Mello Koch and Ramgoolam \cite{Koch:2010zza} (see also the earlier work \cite{Itzykson:1990zb, DiFrancesco:1992cn, itzbauer}) that the combinatorics for computing Gaussian correlators 
$\langle \prod_{i=1}^n {\rm Tr} M^{2k_i} \rangle_{g}$ is suggestive of a sum over branched covers from a genus $g$ worldsheet to a target 
${\mathbb P}^1$, with {\it three} branchpoints\footnote{The authors of \cite{deMelloKoch:2012tb} have also proposed an intriguing interpretation of this in terms of a three dimensional target space: a sphere with three holes times $S^1$.}. Such holomorphic maps are known as Belyi maps. This interpretation of the combinatorics was given a concrete realization in \cite{Gopakumar:2011ev} (with a crucial modification, though, which restricts one for the time being to planar worldsheets) in terms of a specific prescription to glue Feynman diagrams. This also enables one to give an explicit form for the Belyi maps in question using results of \cite{mulpenk1}. 
In essence, this prescription to glue Feynman diagrams is a special application of the general approach to open-closed string duality put forward in \cite{Gopakumar:2003ns, Gopakumar:2004qb, Gopakumar:2004ys, Gopakumar:2005fx}\footnote{For further elaborations on this proposal and related aspects see \cite{Furuuchi:2005qm, Aharony:2006th, David:2006qc, 
Yaakov:2006ce, Aharony:2007fs, Aharony:2007rq, David:2008iz, Brown:2010pb}.} as adapted to the matrix model case by Razamat \cite{Razamat:2008zr, Razamat:2009mc}. 

Furthermore, the target space ${\mathbb P}^1$ was identified \cite{Gopakumar:2011ev} with the riemann surface canonically associated to the Gaussian matrix model. This is the master field geometry which captures the complexified eigenvalue distribution (``Wigner semicircle"). 
One then sees a skeletal version of AdS/CFT in the scattering on the target ${\mathbb P}^1$. 
Finally, since the conventional worldsheet theory describing holomorphic maps to a target is the A-model topological string theory, an attempt was made to directly 
link this to the Gaussian matrix model. 
As described above, some limited evidence was found in favor of this last proposal. 

The additional evidence we provide here is much stronger and can be summarized as follows.
On the matrix model side one can explicitly compute the planar correlators ($n \geq 2$)
\be{matcorr}
\langle \frac{1}{k_1}{\rm Tr}M^{2k_1}\frac{1}{k_2}{\rm Tr}M^{2k_2}\ldots \frac{1}{k_n}{\rm Tr}M^{2k_n} \rangle_{conn} = \frac{(d-1)!}{(d-n+2)!}\prod_{i=1}^{n}\frac{(2k_i) !}{k_i ! k_i! }
\ee
where $d=\sum_{i=1}^nk_i$. 
We will then compare with the genus zero topological string correlators, of gravitational descendants of the kahler class operator, 
which turn out to be ($n \geq 2$) 
\be{topcorr}
\langle (2\sigma_{2k_1-1})(2\sigma_{2k_2-1})\prod_{i=3}^n \sigma_{2k_i}
\rangle_{g=0} = d^{n-3}\prod_{i=1}^{n}\frac{(2k_i) !}{k_i ! k_i!}.
\ee
Firstly note that the RHS is symmetrical in the $k_i$ despite the LHS not being manifestly symmetric ($k_1,k_2$ singled out). We can view these two vertex operators as being at some fixed positions e.g. $z=1$ and $z=\infty$ on the worldsheet. Secondly, we note that \eq{matcorr} and \eq{topcorr} exactly match for $n=2,3$ for any $k_i$. As we will also see, this matching of two and three point functions holds for a general single trace operator ${\rm Tr}M^{p}$ i.e. we do not have to restrict to even powers. 

In a nontrivial AdS/CFT duality, typically, one compares two and three point functions on both sides since higher point functions are determined in terms of these through  factorisation. In that sense we have made a successful comparison of both sides. But here we also have the luxury to see how things might work for higher point functions (at the planar level).

We note that the higher $n$-point functions of the matrix model are close to the string answer though not exactly the same. In fact, for $n > 4$ the prefactor
\be{rel1}
\frac{(d-1)!}{(d-n+2)!} = d^{n-3} - \frac{(n-2)(n-3)}{2}d^{n-2} - \ldots
\ee
Thus the leading piece (for large degree $d$) is indeed that of the string correlator but there also are some corrections which are subleading. Can we interpret these subleading correction terms as well? 

It turns out that there is a natural interpretation. We have the inverse relation
\be{stirling}
d^{n-3} = \sum_{m=3}^{n}\tilde{S}^{(m-2)}_{n-2} \frac{(d-1)!}{(d-m+2)!}
\ee
where the (positive) coefficients $\tilde{S}^{(m-2)}_{n-2}$ turn out to 
count the number of ways in which to partition $(n-2)$ elements into $(m-2)$ non-empty 
subsets (see Sec. 9.74 of \cite{gradryzh}, for example). If we consider the 
$(n-2)$ vertex operators $\sigma_{2k_3}, \ldots \sigma_{2k_n}$ then  
their insertions on the worldsheet can come close to each other. Let's say we have $(m-2)$ such groups of these operators where $(m-2)$ can vary between $(n-2)$
(all of them separate) and $1$ (all operators together at $z=0$, say). $\tilde{S}^{(m-2)}_{n-2}$ simply counts the number of such groupings. 

The interpretation of \eq{stirling} is then that there are contributions from ``contact terms" in the topological string theory when these operators collide (see for e.g. \cite{Verlinde:1990ku}) which must correspond to  lower $m$-point function matrix correlators. Thus if 
$\sigma_{2k_3}$ and $\sigma_{2k_4}$ ``come together''\footnote{Strictly speaking,
points on the worldsheet never come together, one is simply going to a boundary in moduli space, where a sphere pinches off. Nevertheless we will use this loose terminology.} then  by the interpretation of \cite{Gopakumar:2011ev} two ramification points on the worldsheet coincide. On the matrix model side this is possible only if we replace ${\rm Tr}M^{2k_3}{\rm Tr}M^{2k_4} \rightarrow {\rm Tr}M^{2(k_3+k_4)}$ giving rise to an $(n-1)$-point correlator. Note that this is necessitated by the fact that there is no OPE on the matrix model side corresponding to bringing the matrix operators together.
As a result, for four or higher point correlators we have to {\it separately} consider, in the matrix model, the contribution of these contact terms in the topological string correlators. The combinatorial coefficients in \eq{stirling} account for these additional contributions and thus gives a natural way to connect the string answers to that of the matrix model. Note that the operators $\sigma_{2k_1-1}, \sigma_{2k_2-1}$ appear in \eq{topcorr} on a different footing from the $\sigma_{2k}$.

We will therefore also discuss a closely related set of correlators, where we will find a similar relation, but which now treats all the $k_i$ on the same footing. This is in terms of the $(n+2)$-point string correlator $\langle P^2 \prod_{i=1}^n \sigma_{2k_i}\rangle_{g=0}$. The dual matrix correlator must now involve an operator corresponding to the puncture operator $P$. We identify this operator to be ${\rm Lim}_{p\rightarrow 0}\frac{1}{p}{\rm Tr}M^{2p} \sim 2{\rm Tr}\ln M$. We note that a similar identification was proposed by Eguchi and Yang \cite{Eguchi:1994in} in their (closely related) matrix model for the A-model on
${\mathbb P}^1$. We will see that we can compute correlators of this operator both by analytic continuation ($p \rightarrow 0$) as well as directly using standard matrix model technology. Once again using \eq{stirling} we obtain a relation between matrix and string correlators which is, in addition, symmetric in all the $k_i$.

In sections 2 and 3 we describe the matrix and string results respectively. Sec. 4 elaborates on the above comparison while Sec. 5 closes with general remarks. Appendices A and B give more details of the calculation of correlators.


\section{Matrix Correlators}

We will (mostly) consider the subset of matrix correlators in the Gaussian matrix model with even powers ${\rm Tr}M^{2p}$ i.e. $\langle \prod_{i=1}^n {\rm Tr} M^{2k_i} \rangle_{conn}$ and in the planar limit \cite{Brezin:1977sv}. These can be obtained from the generating function
\be{}
Z[t] = \int[dM]_{N\times N}e^{-\frac{1}{2}N{\rm Tr}M^2+\sum_{k} t_{p}N{\rm Tr}M^{2p}}
\ee
by differentiating appropriately with respect to the $t_{k_i}$, taking the logarithm and finally the large $N$ limit. 
This can be done in a variety of ways. One straightforward approach is to use the technique of orthogonal polynomials to write down a general form for the answer. After taking the logarithm to obtain the connected piece we can then take its large $N$ limit.

The generating function can be expressed in terms of the integral over eigenvalues \(\lambda_{i}\) of $M$ (see for e.g. \cite{Ginsparg:1994})
\bea{partf1}
 Z[t] &=& \int\prod_{i}d\lambda_{i}\Delta^2(\lambda)e^{-N\sum_{i}{\rm V}(\lambda)}
\cr \rm{V}(\lambda) &=&\frac{1}{2}\lambda^2-\sum_{p} t_{p}\lambda^{2p}\cr
  \Delta(\lambda) &=& {\rm det}\lambda_{i}^{j-1}.
\eea
We express the Vandermonde determinant $\Delta(\lambda)$ in terms of orthogonal polynomials
\(P_{m}(\lambda)\), which satisfy the orthogonality relation
\be{orth1}
\int d\lambda P_{m}(\lambda)P_{l}(\lambda)e^{-N\rm{V}(\lambda)} = h_{m}\delta_{ml}.
\ee
Then the generating function reads

 \be{partfr1}
 Z = N!h_0^N \prod_{j=0}^{N-1} R_j^{N-j}
\ee
where \(R_{m} = \frac{h_{m}}{h_{m-1}}\). Using the recursion relations of $P_m(\lambda)$  we can derive
the equation which determines $R_{m}$
\be{nlrela1}
 R_m(t)[1-\sum_{k=1}^{\infty}  \frac{(2k)!}{k!(k-1)!}t_{k}R^{k-1}_m(t)]=\frac{m}{N}.
\ee
In the planar limit the rescaled index $m/N$ becomes a continuous variable $y$ that take values in $(0,1)$ and $R_m(t)$ becomes a continuous function
$R(t,y)$. Then the generating function for connected correlators reduces to a simple one-dimensional integral 
\bea{genr1a}
 G(t) &=& \lim_{N \to \infty}\frac{1}{N^2}{\rm ln}(\frac{Z[t_{k}]}{Z[0]}) 
 \cr &=& \int_{0}^{1}dy(1-y){\rm ln}(\frac{R(t,y)}{y}).
 \eea
We solve for $R(t,y)$ from the continuum version of \eq{nlrela1} and 
using \eq{genr1a} we obtain $G(t)$.  
We can then extract the connected correlators from $G(t)$. Some of the steps are shown are shown in Appendix A. 
The final answer is (for $n \geq 2$)
\be{matcorr2}
\langle \frac{1}{k_1}{\rm Tr}M^{2k_1}\frac{1}{k_2}{\rm Tr}M^{2k_2}\ldots \frac{1}{k_n}{\rm Tr}M^{2k_n} \rangle_{conn} = \frac{(d-1)!}{(d-n+2)!}\prod_{i=1}^{n}\frac{(2k_i) !}{k_i ! k_i! }.
\ee
Here $d= \sum_i k_i$.
This agrees with the enumeration of graphs in \cite{tutte} (see also \cite{Ambjrn2d}).
The nontrivial $d$ dependence shows that the answers are not, in general, factorised.
Note that $d$ also has an interpretation as the degree of the Belyi map that contributes to this correlator \cite{Gopakumar:2011ev}. 
It turns out that one can also evaluate the correlators with two odd powers (correlators with one odd power vanish) using the results of \cite{tutte}
\be{matcorr3}
\langle \frac{1}{2k_1+1}{\rm Tr}M^{2k_1+1}\frac{1}{2k_2+1}{\rm Tr}M^{2k_2+1}\prod_{i=3}^n\frac{1}{k_n}{\rm Tr}M^{2k_n} \rangle_{conn} = \frac{d!}{(d-n+3)!}\prod_{i=1}^{n}\frac{(2k_i) !}{k_i ! k_i! }.
\ee
with $d= \sum_i k_i$.

From these results we see that, in particular the two point function is given by
\be{mat2pt}
\langle  {1\over k_1}{\rm Tr } M^{2k_1} {1\over k_2}{\rm Tr } M^{2k_2} \rangle_{conn} = \frac{1}{k_1+k_2}\frac{(2k_1)!}{(k_1!)^2}\frac{ (2k_2)!}
{(k_2!)^2}.
\ee
which agrees with the calculation in \cite{Gopakumar:2011ev} since $d= k_1+k_2$ in this case. 
We also have 
\be{mat2pt2}
\langle  {1\over 2k_1+1}{\rm Tr } M^{2k_1+1} {1\over 2k_2+1}{\rm Tr } M^{2k_2+1} \rangle_{conn} = \frac{1}{k_1+k_2+1}\frac{(2k_1)!}{(k_1!)^2}\frac{ (2k_2)!}
{(k_2!)^2}.
\ee

Interestingly the three point function is the only one which is factorised. We have the non-vanishing ones to be 
\be{mat3pt}
 \langle  {1\over k_1}{\rm Tr } M^{2k_1} {1\over k_2}{\rm Tr } M^{2k_2} {1\over k_3}{\rm Tr } M^{2k_3}\rangle_{conn} = \frac{(2k_1)!}{(k_1!)^2}\frac{ (2k_2)!}
{(k_2!)^2}\frac{(2k_3)!}{(k_3!)^2}. 
\ee
and
\be{mat3pt2}
 \langle  {1\over 2k_1+1}{\rm Tr } M^{2k_1+1} {1\over 2k_2+1}{\rm Tr } M^{2k_2+1} {1\over k_3}{\rm Tr } M^{2k_3}\rangle_{conn} = \frac{(2k_1)!}{(k_1!)^2}\frac{ (2k_2)!}
{(k_2!)^2}\frac{(2k_3)!}{(k_3!)^2}. 
\ee

As mentioned in the introduction we will also compare correlators with (two) insertions of the operator ${\rm Tr}\ln M$. These can also be explicitly evaluated as we show in Appendix A.
\bea{lncorr}
\langle ({\rm Tr}\ln M)^2 \prod_{i=1}^n \frac{1}{k_i}{\rm Tr} M^{2k_i} \rangle_{conn}
&=& \frac{(d-n+2)(d-n+1)}{4} \langle  \prod_{i=1}^n {\rm Tr} M^{2k_i} \rangle_{conn} \cr
& =& \frac{(d-1)!}{4(d-n)!}\prod_{i=1}^{n}\frac{(2k_i) !}{k_i ! k_i! }.
\eea
A heuristic way to obtain this answer is to consider the $(n+2)$ point function
\be{lnheur}
{\rm Lim}_{\epsilon_{1,2}\rightarrow 0}
\langle   \frac{1}{\epsilon_1}{\rm Tr} M^{2\epsilon_1}
\frac{1}{\epsilon_2}{\rm Tr} M^{2\epsilon_2}\prod_{i=1}^n \frac{1}{k_i}{\rm Tr} M^{2k_i} \rangle_{conn}
= \frac{(d-1)!}{(d-n)!}\prod_{i=1}^{n}\frac{(2k_i) !}{k_i ! k_i! }.
\ee
Thus analytically continuing in $\epsilon_i$ and using ${\rm Lim}_{p\rightarrow 0}\frac{1}{p}{\rm Tr}M^{2p} \sim 2{\rm Tr}\ln M$ (the constant piece does not contribute to the connected correlator) we obtain the answer in \eq{lncorr}.

In the description in terms of Belyi maps given in \cite{Gopakumar:2011ev} 
both sets of correlators in \eq{matcorr2} and \eq{lncorr} get contributions only from maps of degree $d=\sum_i k_i$. For the latter, we can understand this using the description of the logarithmic operator as in \eq{lnheur}.

\section{String correlators}

Correlators in the A-model topological string on ${\mathbb P}^1$ are determined by recursion relations. The main relations are summarized in, for instance, \cite{Eguchi:1996tg}. 
The observables in the theory are the puncture operator ${\cal V}_1= P$, the operator corresponding to the Kahler class ${\cal V}_2=Q$ and their 
gravitational descendants $\sigma_n(P), \sigma_n(Q)$ (for $n > 0$). The recursion relation we will mostly employ is 

\be{trr1}
 \langle \sigma_{n}({\cal V}_{\gamma})XY\rangle_{g=0} = n \langle \sigma_{n-1}({\cal V}_{\gamma}){\cal V}_{\alpha}\rangle\eta^{\alpha\beta}\langle{\cal V}_{\beta}XY\rangle_{g=0}
\ee
where $X,Y $ are arbitrary observables and this holds in the so-called large phase space i.e. with arbitrary backgrounds for the descendants turned on as well. Therefore this enables one to express a general $n$-point function in terms of less complicated ones. 
In this paper we will restrict ourselves to correlators involving the puncture operator $P$ as well as  $\sigma_n(Q)$ which we will henceforth denote by $\sigma_n$ (for $n> 0$) without hopefully causing confusion.


Using these and other recursion relations we find for $n\geq 2$ (see Appendix B for details)
\be{strnpt}
\langle (2\sigma_{2k_1-1})(2\sigma_{2k_2-1})\prod_{i=3}^n \sigma_{2k_i}
\rangle_{g=0} = d^{n-3}\prod_{i=1}^{n}\frac{(2k_i) !}{k_i ! k_i!}.
\ee
Here, as before $d=\sum_i k_i$ and is also the degree of the holomorphic map which contributes to the above correlator, as can be seen from the selection rule given in \eq{gconl}. We note that though the left hand side of \eq{strnpt} is not manifestly symmetric in all the $k_i$, the answer on the RHS is nevertheless so. 
We also record the answer for the correlator with all even powers 
\be{strnpt3}
\langle \prod_{i=1}^{n}\sigma_{2k_i}\rangle_{g=0} =  (d+1)^{n-3}\prod_{i=1}^{n}\frac{(2k_i)!}{(k_i!)^2}. 
\ee
Here we retain the notation $d=\sum_i k_i$ but caution that the degree of the map which contributes to this correlator is actually $(d+1)$. 

We will also use the related result that (for $n\geq 1$)
\be{strnpt2}
\langle P^2\prod_{i=1}^{n}\sigma_{2k_i}\rangle_{g=0} = d^{n-1}\prod_{i=1}^{n}\frac{(2k_i) !}{k_i ! k_i!}. 
\ee

The selection rule \eq{gconl} for the correlator in \eq{strnpt2}  shows that the contributions come only from holomorphic maps of degree $d =\sum_i k_i$.

\section{Comparison}

We can now compare the results on both sides. 

\subsection{Two and Three Point functions}

We firstly note that 
the two and three point functions agree for arbitrary $k_i$. We see from \eq{mat2pt} and \eq{strnpt} for $n=2$ that
\bea{2ptcomp}
\langle  {1\over k_1}{\rm Tr } M^{2k_1} {1\over k_2}{\rm Tr } M^{2k_2} \rangle_{conn}  &=& \frac{1}{k_1+k_2}\frac{(2k_1)!}{(k_1!)^2}\frac{(2k_2)!}
{(k_2!)^2} \cr
& & \cr
& =& \langle (2\sigma_{2k_1-1})(2\sigma_{2k_2-1})
\rangle_{g=0}.
\eea
as well as \eq{mat2pt2} and \eq{strnpt2} for $n=2$ that
\bea{2ptcomp2}
\langle  {1\over 2k_1+1}{\rm Tr } M^{2k_1+1} {1\over 2k_2+1}{\rm Tr } M^{2k_2+1} \rangle_{conn}  &=& \frac{1}{k_1+k_2+1}\frac{(2k_1)!}{(k_1!)^2}\frac{(2k_2)!}
{(k_2!)^2} \cr
& & \cr
& =& \langle \sigma_{2k_1}\sigma_{2k_2}
\rangle_{g=0}.
\eea

Similarly, from \eq{mat3pt} and \eq{mat3pt2} together with \eq{strnpt} and \eq{strnpt3} for $n=3$ we have  
\bea{3ptcomp}
\langle  {1\over k_1}{\rm Tr } M^{2k_1} {1\over k_2}{\rm Tr } M^{2k_2} {1\over k_3}{\rm Tr } M^{2k_3}\rangle_{conn}  &=& \frac{(2k_1)!}{(k_1!)^2}\frac{ (2k_2)!}
{(k_2!)^2}\frac{(2k_3)!}{(k_3!)^2} \cr
&& \cr
& =& \langle (2\sigma_{2k_1-1})(2\sigma_{2k_2-1})\sigma_{2k_3}
\rangle_{g=0}.
\eea
and
\bea{3ptcomp2}
\langle  {1\over 2k_1+1}{\rm Tr } M^{2k_1+1} {1\over 2k_2+1}{\rm Tr } M^{2k_2+1} {1\over k_3}{\rm Tr } M^{2k_3}\rangle_{conn}  &=& \frac{(2k_1)!}{(k_1!)^2}\frac{ (2k_2)!}
{(k_2!)^2}\frac{(2k_3)!}{(k_3!)^2} \cr
&& \cr
& =& \langle \sigma_{2k_1}\sigma_{2k_2}\sigma_{2k_3}
\rangle_{g=0}.
\eea

As mentioned in the introduction, this in itself is a fairly good check that things are on the right track. 

\subsection{General correlators}

We can however go on to compare the general $n$ point function. The answers are given in \eq{matcorr} and \eq{topcorr} (or \eq{matcorr2} and \eq{strnpt}). We see that for $n \geq 4$ they are not quite identical. But this near agreement is quite remarkable in itself for a couple of reasons. Firstly, apart from the individual factors $\frac{(2k_i)!}{k_i ! k_i!}$ which is a dependence that can be absorbed into a redefinition of the individual operators, both sets of correlators could have depended on arbitrary symmetric functions of the $k_i$'s in a complicated way. The fact that both sides should, a priori, have depended only on the particular symmetric 
combination $d=\sum_i k_i$ is not obvious. Secondly, as seen in \eq{rel1}, 
the combination $\frac{(d-1)!}{(d-n+2)!}$ is a polynomial in $d$ of degree $(n-3)$ with the leading term the same as the string answer. Thus there is exact agreement in the large $d$ (or large $k_i$) regime which is some kind of BMN like limit. The large $k_i$ regime is where the Feynman diagrams are dominated by graphs with a large number of edges and faces. From the point of view of the moduli space one is getting contributions from many more points on the moduli space. Effectively one will have a continuum description of moduli space \cite{Razamat:2008zr}\footnote{See \cite{Levin:1990fa, Smit:1990nv} for a discussion of the subtleties in how equilateral triangulations capture the continuum measure on moduli space.}.  

We will now elaborate on the relation between the general matrix $n$-point correlator and that of the string theory which was sketched in the introduction. Firstly we will normalize the operators on both sides so as to get rid of the factors $\frac{(2k_i) !}{k_i ! k_i!}$. Namely, we define
\be{redef}
\tilde{\sigma}_{2k} = \frac{(k!)^2}{(2k)!}\sigma_k ; \qquad \qquad {\cal O}_{2k} =\frac{1}{k} \frac{(k!)^2}{(2k)!}
{\rm Tr } M^{2k}.
\ee
Then we have to compare
\be{redefstr}
\langle (2\tilde{\sigma}_{2k_1-1})(2\tilde{\sigma}_{2k_2-1})\prod_{i=3}^n \tilde{\sigma}_{2k_i}\rangle_{g=0} = d^{n-3}
\ee 
with 
\be{}
\langle \prod_{i=1}^n {\cal O}_{2k_i} \rangle_{conn} = \frac{(d-1)!}{(d-n+2)!}.
\ee    

Using the relation \eq{stirling} we can write
\be{rel2}
\langle (2\tilde{\sigma}_{2k_1-1})(2\tilde{\sigma}_{2k_2-1})\prod_{i=3}^n \tilde{\sigma}_{2k_i}\rangle_{g=0} = \sum_{m=3}^{n}\tilde{S}^{(m-2)}_{n-2} \langle {\cal O}_{2k_1}{\cal O}_{2k_2}\prod_{j=1}^{m-2} {\cal O}_{2\mu_j} \rangle_{conn}.
\ee
Here $\mu_j = \sum_{r\in R_j}k_{r}$ with $(j=1 \ldots (m-2))$ where $\{ R_j \}$ are $(m-2)$ different non-empty groupings of the $(n-2)$ integers $(3, 4, \ldots n)$.  Thus the ${\cal O}_{2\mu_j}$ are essentially operators of the form ${\rm Tr}M^{(\sum_{R_j} 2k_{r_j})}$ over different groupings of the set $(k_3, \ldots k_n)$.  

The stirling number of the second kind, $\tilde{S}^{(m-2)}_{n-2}$, which appears in \eq{stirling}
precisely counts the number of ways in which we can partition $(3, \ldots n)$ 
into the $(m-2)$ sets $R_j$ such that each $R_j$ contains at least one integer.  Note that since the matrix correlator only depends on $\sum_i k_i$, we have 
\be{}
\langle {\cal O}_{2k_1}{\cal O}_{2k_2}\prod_{j=1}^{m-2} {\cal O}_{2\mu_j} \rangle_{conn} = 
\frac{(d-1)!}{(d-m+2)!}
\ee 
independent of the partitioning (i.e. the $\mu_j$). We only need that $k_1+k_2 + \sum_j \mu_j =\sum_i k_i =d$. This, in turn, is what enables us to write \eq{rel2} or equivalently
\be{rel3}
\langle (2\tilde{\sigma}_{2k_1-1})(2\tilde{\sigma}_{2k_2-1})\prod_{i=3}^n \tilde{\sigma}_{2k_i}\rangle_{g=0} =  \sum_{m=3}^{n} \sum_{partitions \{R_j \}}\langle {\cal O}_{2k_1}{\cal O}_{2k_2}\prod_{j=1}^{m-2} {\cal O}_{2\mu_j} \rangle_{conn}.
\ee

Recall that the interpretation of ${\rm Tr}M^{2p}$ in \cite{Gopakumar:2011ev} was that it created a ramification point of order $p$ on the worldsheet. The map to the target ${\mathbb P}^1$ therefore locally looks like $X(z) = (z-z_1)^p$ near this vertex operator insertion.
In the string theory we can have two ramification points with behavior $(z-z_1)^{p_1}$ and $(z-z_2)^{p_2}$ 
coming together when $z_1 \rightarrow z_2$ to create a ramification point of order $(p_1+p_2)$. However, in the matrix model, unlike in a QFT, 
we do not have an OPE of the two corresponding operators 
${\rm Tr}M^{2p_1}$ and ${\rm Tr}M^{2p_2}$ giving something like ${\rm Tr}M^{(2p_1+2p_2)}$.
We have to put in the contribution to the string correlator, from the collision of ramification points, {\it separately} on the matrix model side. They are not contained in the original $n$-point correlator
$\langle \prod_i^n {\rm Tr} M^{2k_i} \rangle_{conn}$.  
But we now see that we can interpret the different terms on the right hand side of \eq{rel3} as these additional contributions. Thus, in addition to the $n$-point matrix correlator (corresponding to $m=n$ or the partition where each of the $R_j$ contain exactly one integer) we also have the lower point functions all the way unto a three point function (where we have 
all of $(k_3, \ldots k_n)$ come together). In general, through these lower point functions we include all the contributions where various of the $\sigma_{2k_i}$ (for $i=3,\ldots n)$ come together in different groupings. 

Note that we do not have any contribution corresponding to bringing any of the $\sigma_{2k_i}$ near either of  $\sigma_{2k_1-1}$ or  $\sigma_{2k_2-1}$. It remains to be understood from the string point of view why this is the case. Admittedly, this is somewhat unsatisfactory in that it treats the $k_i$ in an asymmetric way despite the RHS being symmetric\footnote{One can make a similar comparison  for the string correlator \eq{strnpt3} (which is symmetric in the $k_i$)  with the matrix correlator \eq{matcorr3} (which singles out $k_1, k_2$).
\be{rel4}
\langle \tilde{\sigma}_{2k_1}\tilde{\sigma}_{2k_2}\prod_{i=3}^n \tilde{\sigma}_{2k_i}\rangle_{g=0} =  \sum_{m=3}^{n} \sum_{partitions \{R_j \}}\langle {\cal O}_{2k_1+1}{\cal O}_{2k_2+1}\prod_{j=1}^{m-2} {\cal O}_{2\mu_j} \rangle_{conn}.
\ee
}. We will therefore remedy this by looking at an alternate set of correlators where the symmetry is manifest. However, this will be at the expense of introducing additional operators of the form ${\rm Tr}\ln M$. 

On the string theory side the correlators we consider are the ones in \eq{strnpt2} rewritten using the redefined operators in \eq{redef} as
\be{strnpt21}
\langle P^2\prod_{i=1}^{n}\tilde{\sigma}_{2k_i}(Q)\rangle_{g=0} = d^{n-1}. 
\ee
With the identification $P \leftrightarrow 2{\rm Tr}\ln M \equiv {\cal P}$ we compare with the matrix correlator \eq{lncorr}
\be{matnpt2}
\langle {\cal P}^2 \prod_i^n {\cal O}_{2k_i} \rangle_{conn} = \frac{(d-1)!}{(d-n)!}.
\ee
We see that there is a mismatch for $n\geq 2$. But once again we have a natural relation between the two sides. We use
\be{stirl2}
d^{n-1} = \sum_{m=1}^{n}\tilde{S}^{(m)}_{n} \frac{(d-1)!}{(d-m)!}
\ee
to write
\bea{relpun}
\langle P^2\prod_{i=1}^n \tilde{\sigma}_{2k_i}\rangle_{g=0} & = & 
\sum_{m=1}^{n}\tilde{S}^{(m)}_{n} \langle {\cal P}^2 \prod_{j=1}^{m} {\cal O}_{2\mu_j} \rangle_{conn} \cr
& = &  \sum_{m=1}^{n} \sum_{partitions \{R_j \}}\langle {\cal P}^2 \prod_{j=1}^{m} {\cal O}_{2\mu_j} \rangle_{conn}.
\eea
The partitions $\{ R_j \}$ (with $j = 1\ldots m$) are the $m$ different non-empty groupings of the $n$ integers $(1,2, \ldots n)$. Thus the ${\cal O}_{2\mu_j}$ are essentially operators of the form ${\rm Tr}M^{(\sum_{R_j} 2k_{r_j})}$ over different groupings of the set $(k_1, k_2, \ldots k_n)$. 

The interpretation of the RHS in \eq{relpun} is therefore similar to before. Now we have all the $n$ operators $\sigma_{2k_i}$ on the same footing on the LHS 
and so they can all come close to each other leading to the merging of ramification points. The RHS counts the contributions of these separate groupings from the matrix model side where we have lower point functions involving the merged operators ${\rm Tr}M^{(\sum_{R_j} 2k_{r_j})}$. Note that the puncture operator 
$P \leftrightarrow {\rm lim}_{\epsilon \rightarrow 0} {\rm Tr} M^{2\epsilon}$ does not create any branching. In many ways this is a much more neat picture for the correspondence between the correlators in the matrix model and the dual string theory.

\section{Discussion}

What we have learnt from this comparison of correlators in the topological string theory on ${\mathbb P}^1$ with those of the Gaussian matrix model is that they are not identically the same except for the two and three point functions. In fact, in hindsight, we see that for four or higher point functions there was no reason to have expected them to be the same since the 
matrix correlator does not allow for the possibility of bringing operators together on the 
worldsheet. Instead, the relation between the two sets of correlators is one in which we add in, on the matrix model side, the separate contributions from the fusing of two matrix operators.
On the topological string theory side it would be good to understand more explicitly these contact terms. Note that in the usual approach, such as \cite{Verlinde:1990ku}, the worldsheet is in a gauge where all the curvature is concentrated at the location of the vertex operators. 
However, as remarked in \cite{Gopakumar:2011ev}, the matrix model naturally gives rise to a ``strebel gauge" on the world sheet where the curvature is localized not just at the location of the insertions but also at the interaction vertices. This is likely to affect the contact term contributions and it would be good to see if it exactly matches what we find here. 

We have made a comparison of the general set of even power matrix correlators which is the sector in which the answer is easy to obtain  in closed form. It would be good to extend this to correlators involving an arbitrary number of odd powers as well. This seems to be harder to do explicitly. Another direction to extend the checks is to consider not just the so-called stationary sector of the topological string theory involving the gravitational descendants 
$\sigma_n(Q)$ but also the $\sigma_n(P)$. Here we have looked at correlators with insertion of the puncture operator insertion and seen that they correspond to insertions of ${\rm Tr ln}M$. It is tempting to guess that $\sigma_n(P) \leftrightarrow {\rm Tr}(M^n{\rm ln}M)$ following 
\cite{Eguchi:1994in}. However, as discussed in \cite{Gopakumar:2011ev}, there are important differences between the current proposal and the Eguchi-Yang model, 
which is presumably related to 
the  different relation, proposed here, between the general $n$-point correlators on both sides.  

It is important to extend the relation between both sides beyond the planar/genus zero case. This would require taking into account effects of mixing of single and double trace operators and hence the correspondence between string vertex operators and matrix gauge invariant operators will acquire $\frac{1}{N}$ corrections. 
Another interesting generalization would involve the gaussian {\it normal} matrix model for which there is a proposed dual \cite{Itzhaki:2004te}. It would be nice to make some contact between topological strings, such as the one described here, with the imaginary Liouville backgrounds proposed in \cite{Itzhaki:2004te}\footnote{We thank J. McGreevy for drawing our attention to this work and for comments on related issues.}.
We leave these explorations for the future. 

Finally, the results described here, in addition to their value as a toy model of AdS/CFT, may also be significant in the canonical gauge-string duality between ${\cal N}=4$ super Yang-Mills theory and the string theory on $AdS_5\times S^5$. The localization arguments of \cite{Pestun:2007rz} have shown how the half BPS Wilson loops in the gauge theory reduce to a Gaussian matrix integral. Given the duality elucidated here, one might hazard the guess that there is a corresponding localization of the  $AdS_5\times S^5$ string theory in the half BPS sector which reduces the string sigma model to the A-model topological string theory on ${\mathbb P}^1$. We hope to report on this soon\footnote{See \cite{Bonelli:2008rv} for a proposed relation between the half BPS Wilson loops and a topological sector of the sigma model on $AdS_5\times S^5$.}.

\bigskip

{\bf Acknowledgements:} We would like to thank Shailesh Lal, Sanjaye Ramgoolam, Shlomo Razamat and Ashoke Sen for discussions. 
R.G.'s work was partly supported by a 
Swarnajayanthi Fellowship of the Dept.\ of Science and Technology, Govt.\ of India.  
Both of us acknowledge our debt to the Indian public which has generously supported the basic sciences.

\section*{Appendix} 
\appendix

\section{Correlators in the Gaussian Matrix Model}

 The generating function for arbitrary $n$ point correlators \( \langle \prod_{i}^{n}{\rm Tr}M^{2k_{i}} \rangle\)
of the Gaussian matrix model is
\be{partitionf}
 Z[t] = \int[dM]_{N\times N}e^{-\frac{1}{2}N{\rm Tr}M^2+\sum_{k} t_{k}N{\rm Tr}M^{2k}}.
\ee 
Since \(M\) is a hermitian matrix it is possible to parametrize this in terms
of a unitary matrix \(U\) and eigenvalues \(\lambda_{k}\). i.e; we can write \(M = U^{\dagger}\Lambda U\),
where \(\Lambda = diag(\lambda_{1},...,\lambda_{N})\). Then the generating function can be expressed
in terms of an integral over eigenvalues \(\lambda_{k}\) and evaluated using orthogonal polynomials as discussed in the text. 

We find the final answer to be given as in \eq{partfr1}
\be{partfr}
 Z = N!\prod_{j=0}^{N-1} h_{j} = N!h_0^N \prod_{j=0}^{N-1} R_j^{N-j}
\ee
\\
where \(R_{m} = \frac{h_{m}}{h_{m-1}}\). Thus the calculation of the generating function reduces to calculating \(h_{j}\) or equivalently $R_j$.
To determine these we need to use the recursion formulae for the \(P_{m}\).
The orthogonal polynomials satisfy the usual  three term recursion formula
\\
\be{recur}
 \lambda P_{m}(\lambda) = P_{m+1}(\lambda)+S_{m}P_{m}(\lambda)+R_{m}P_{m-1}(\lambda).
\ee
\\
 This can be derived by observing that l.h.s of \eq{orth1} is symmetric under the exchange of 
 $l$ and $m$.
For even potentials $V(\lambda)$, where $P_m(-\lambda) = (-1)^mP_m(-\lambda)$, we have \(S_{m}=0\).
On the other hand, we obtain a non-linear recursion formula by looking at \(\int d\lambda \lambda P_{m}^{\prime}(\lambda)P_{m}(\lambda)e^{-N\rm{V}(\lambda)} \). This gives
\\
\be{nlrecur}
 NR_{m}\int d\lambda P_{m}(\lambda)P_{m-1}(\lambda)V'(\lambda)e^{-N\rm{V}(\lambda)} = mh_{m}.
\ee
\\

We can now study the large $N$ limit. Assume that $n$ is an integer of order $N$ and $k$ is an integer of order one. Then we can safely assume to leading order

\be{largn}
 R_{m-k}=R_{m-k+1}=...=R_{m-1}=R_m.
\ee
\\
Using repeatedly the orthogonality property and integrating by parts we get
\\
\be{orel}
 \int d\lambda \lambda^{2k-1} P_{m}(\lambda)P_{m-1}(\lambda)e^{-N\rm{V}(\lambda)} = \frac{(2k)!}{k!(k-1)!}h_{m}R^{k-1}_m(t)
.\ee
\\
By $t$ we mean the collection of couplings $\{t_{k} \}$. Then for \(V(\lambda)\) given in \eq{partf1}, relation \eq{nlrecur} reduces to 

\be{nlrela}
R_m(t)(1-A_m(t))=\frac{m}{N}
\ee
\\
where, \(A_m(t)=\sum_k \frac{(2k)!}{k!(k-1)!}t_{k}R^{k-1}_m(t)\). In the planar limit the rescaled index $m/N$ becomes a
continuous variable $y$ that takes values in $(0,1)$, and $R_m(t)$ and $A_m(t)$ become continuous functions
$R(t,y)$ and $A(t,y)$. We use this to calculate various correlators explicitly.

\subsection{Connected correlator $ \langle \prod_{i=1}^{n}{\rm Tr}M^{2k_{i}} \rangle_{conn}$}

The generating function for connected correlators \( \langle \prod_{i=1}^{n}{\rm Tr}M^{2k_{i}} \rangle\) in the large $N$ limit is 
obtained by taking the continuum limit of \eq{partfr}. In this limit, the free energy (up to an irrelevant additive constant) reduces
to a simple one-dimensional integral: 
\bea{genra}
 G(t) &=& \lim_{N \to \infty}\frac{1}{N^2}{\rm ln}(\frac{Z[t]}{Z[0]}) \cr &=& \int_{0}^{1}dy(1-y){\rm ln}(\frac{R(t,y)}{y})
\cr &=& -\int_{0}^{1}dy(1-y){\rm ln}(1-A(t,y)).
 \eea
Here for moving from the second step to the third we have used the following continuum limit of relation \eq{nlrela} 
\bea{contl}
 \frac{R(t,y)}{y} &=& \frac{1}{1-A(t,y)} ,\cr
 A(t,y) &=& \sum_kA(k)t_{k}R^{k-1}(t,y),
\eea
where $A(k) = \frac{(2k)!}{k!(k-1)!}$. 
The free energy is the generating function of connected correlators. Thus we obtain the required correlator by suitable differentiation
\bea{corr}
 \langle\prod_{i}^{n}{\rm Tr}M^{2k_{i}} \rangle_{conn} &=&\frac{\partial^{n}}{\partial t_{k_{1}}...\partial t_{k_{n}}}G(t_{k})|_{t_{k_{i}}=0}
\cr  & =&\int_{0}^{1}dy(1-y) \Bigl(\sum_{m=1}^{n}(m-1)!\sum_{partitions\{R_j\}}\prod_{j=1}^{m}\prod_{r\in R_j}\partial_{t_{k_{r}}}A\Bigr)|_{t_{k_i}=0}\cr &&
 \eea
where \(A = A(t,y)\) and we use the fact that $A(t=0,y)=0$. 
The notation is as in Sec. 4: we have partitioned the $n$ integers $(1,2,\ldots n)$ into $m$ non-empty groupings $\{R_j \}$ where $j=1\ldots m$.  
Examination of \eq{corr} then reveals the structure of the correlator to be 
\\
\be{corrl}
\langle\prod_{i=1}^{n}{\rm Tr}M^{2k_{i}} \rangle_{conn} = T_{n-1}(k_1,...,k_n)(\prod_{i=1}^{n}A(k_{i}))\int_{0}^{1}dy(1-y)R(t=0,y)^{d-n}
\ee
\\
where $ d = \sum_{i=1}^{n}k_{i}$. We also note from \eq{contl} that $R(t=0,y) = y$. And \(T_{n-1}(k_1,...,k_n)\) is a polynomial of order \(n-1\) in each \(k_{i}\)
which is symmetric in all \(k_{i}\). 
It is a combinatorial challenge to work out the form of the polynomial. Surprisingly the answer is simple and given in \cite{tutte} (see also \cite{Ambjrn2d})
\be{poly}
 T_{n-1}(k_1,...,k_n) = \frac{(d-1)!}{(d-n)!}.
\ee
Hence the connected $n$-point correlator in the large $N$ limit is given by 
\be{correlator}
\langle\prod_{i=1}^{n}{\rm Tr}M^{2k_{i}} \rangle_{conn}= \frac{(d-1)!}{(d-n+2)!}\prod_{i=1}^{n}\frac{(2k_i)!}{k_i!(k_i-1)!}.
\ee

\subsection{Connected correlator $ \langle ({\rm Tr} \ln M)^2\prod_{i=1}^{n}{\rm Tr}M^{2k_{i}} \rangle_{conn}$}

The generating function for the arbitrary $(n+2)$-point correlator $\langle ({\rm Tr}\ln M)^2\prod_{i=1}^{n}{\rm Tr}M^{2k_{i}} \rangle$
 is given by
\be{lnpartf}
 Z[t,\alpha] = \int[dM]_{N\times N}e^{-\frac{1}{2}N{\rm Tr}M^2+\sum_{k} t_{k}N{\rm Tr}M^{2k}+\alpha N{\rm Tr}\ln M^2}
.\ee
We can evaluate the required correlator by following the arguments given in previous section. The potential $V(\lambda)$ is slightly
modified to have a logarithmic term,
\be{lnpot}
 {\rm V}(\lambda) =\frac{1}{2}\lambda^2-\sum_{k} t_{k}\lambda^{2k}+\alpha\ln\lambda^2.
\ee
Here too we can write down a relation like \eq{nlrela}, which we derived by taking the large 
$N$ limit of the relation
 \eq{nlrecur}. 
When $m$ is odd ${\lambda}^{-1} P_m(\lambda)$ is a polynomial and can be expressed in terms of $ P_{m-1}(\lambda)$ and 
 $ P_{m-2}(\lambda)$ using \eq{recur}. Note that we have an even potential so $S_m$ is again zero. Using this we can derive the 
following
\bea{lnorth}
\int d\lambda \lambda^{-1} P_{m}(\lambda)P_{m-1}(\lambda)e^{-N\rm{V}(\lambda)} &=& h_{m-1} \rm{,~~ for~ }  m  \rm{ ~odd} \cr 
&=& 0    \rm{ , ~~~~~~~for~ } m  \rm{ ~even }.
\eea
This leads us to the required relation 
\bea{lnlnrel}
R_m(t,\alpha)[1-\sum_kt_{k}A(k)R_m^{k-1}(t,\alpha) - 2\frac{\alpha}{R_m(t,\alpha)}]&=& \frac{m}{N} \rm{,~~ for~ } m \rm{ ~odd} \cr 
R_m(t,\alpha)[1-\sum_k t_{k}A(k)R_m^{k-1}(t,\alpha)]&=& \frac{m}{N}   \rm{ , ~~for~}  m  \rm{ ~even }.
\eea
For large $N$ we can take a continuum limit as in the previous case (we now take an average of the cases with $m$ even and $m$ odd) and get
\bea{lnrelation}
\frac{R(t,\alpha,y)}{y} &=& \frac{1}{1-A(t,\alpha,y)} \cr
A(t,\alpha,y) &=& \sum_kA(k)t_{k}R^{k-1}(t,\alpha,y)+\frac{\alpha}{R(t,\alpha,y)}. 
\eea
Thus the generating function for the large $N$ connected  correlator  $ \langle ({\rm Tr}\ln M)^2\prod_{i}^{n}{\rm Tr}M^{2k_{i}} \rangle_{conn}$
is
\be{genra}
G(t,\alpha) = -\int_{0}^{1}dy(1-y){\rm ln}[1-A(t,\alpha,y)].
\ee
Now without much difficulty we can extract the correlator from the generating function as we did in the previous case.
\bea{lncorrila}
\langle ({\rm Tr}\ln M)^2\prod_{i=1}^{n}{\rm Tr}M^{2k_{i}} \rangle_{conn} &=&\frac{\partial^{n+2}}{4\partial\alpha^2\partial t_{k_{1}}...\partial t_{k_{n}}}G(t,\alpha)|_{t=0,\alpha =0}
\cr  & =&\frac{\partial^{2}}{4\partial\alpha^2}\langle\prod_{i=1}^{n}{\rm Tr}M^{2k_{i}} \rangle_{conn}|_{\alpha =0}
\cr  &=&\frac{(d-1)!}{4(d-n)!}\prod_{i=1}^{n}A(k_i)\int_{0}^{1}dy(1-y)\frac{\partial^{2}}{\partial\alpha^2}R^{d-n}(t=0,\alpha,y)|_{\alpha =0}.\cr& &
\eea
From \eq{lnrelation} it is clear that $R(t=0,\alpha,y)=y+\alpha$. Thus we get 
\be{lncorrelator}
\langle ({\rm Tr}\ln M)^2\prod_{i=1}^{n}{\rm Tr}M^{2k_{i}} \rangle_{conn}= \frac{(d-1)!}{4(d-n)!}\prod_{i=1}^{n}\frac{(2k_i)!}{k_i!(k_i-1)!}.
\ee

\section{Correlators of the topological A-model string theory on ${\mathbb P}^1$ }

 Physical observables in topological A-model string theory on ${\mathbb P}^1$ arises from the cohomology of the target manifold 
 ${\mathbb P}^1$. They are the puncture operator $P$, Kahler class $Q$  and their graviational descendants $\sigma_{n}(Q)$,
 $\sigma_{n}(P)$ (for $n\geq 1$). The partition function of this topological string theory depends on a set of couplings $\{ t_k, t_k^{\prime} \}$ corresponding to these operators. 
This is the generating function of all the correlators in the theory. 
The genus $g$ correlation functions $\langle\prod_{i=1}^{n}\sigma_{2k_i}({\cal V}_{\alpha_i})\rangle_{g}$ (in the background where all couplings $t_k, t_k^{\prime}$ vanish) receives contributions only from holomorphic maps (from the world sheet to the target ${\mathbb P}^1$)  of degree $d$ satisfying the ghost number conservation law (see for e.g. Eq.(2.24) of \cite{Eguchi:1995er})
 \be{gconl}
 2d+2(g-1) =  \sum_{i=1}^{n}(2k_i+q_{\alpha_i}-1)
 \ee
where $ {\cal V}_{1} = P,  {\cal V}_{2} = Q$ with  $q_{1} = 0, q_{2} = 1$. 
Here we will be considering only a specific set of genus zero correlators namely 
$\langle \sigma_{2k_1-1}(Q)\sigma_{2k_2-1}(Q)\prod_{i=3}^{n}\sigma_{2k_i}(Q)\rangle_{g=0}$ 
and $\langle P^2\prod_{i=1}^{n}\sigma_{2k_i}(Q)\rangle_{g=0}$. This set of correlators defines what is sometimes called the stationary sector of the string theory. For both 
set of correlators the selection rule \eq{gconl} reduces to $2d-2= 2\sum_{i}^n k_i-2$. 
 
To compute these correlators we can use the various recursion relations that they satisfy. 
The important ones are summarized in \cite{Eguchi:1996tg}.
These relations help us to express the $n$-point correlators in terms of low point correlators.
Relevant recursion relations are listed below.
 \\
\\
Topological recursion relation :
\be{trr}
 \langle \sigma_{n}({\cal V}_{\gamma})XY\rangle_{g=0} = n\langle \sigma_{n-1}({\cal V}_{\gamma}){\cal V}_{\alpha}\rangle\eta^{\alpha\beta}\langle{\cal V}_{\beta}XY\rangle_{g=0}
\ee
where $X,Y $ are arbitrary observables and this holds in large phase space - where all the couplings $\{ t_k, t_k^{\prime} \}$ are turned on.
\\
\\
Puncture equation :
\be{punct}
 \langle P\prod_{i=1}^{s}\sigma_{n_i}(Q)\rangle_{g=0} = \sum_{i=1}^{s}n_i\langle \sigma_{n_i-1}(Q)\prod_{j\neq i}\sigma_{n_j}(Q)\rangle_{g=0}.
\ee
\\
Hori's relation \cite{Hori:1994} :
\be{hori}
 \langle Q\prod_{i=1}^{s}\sigma_{n_i}(Q)\rangle_{g=0} = d\langle \prod_{i=1}^{s}\sigma_{n_i}(Q)\rangle_{g=0}
\ee
with $ d =  \frac{1}{2}\sum_{i=1}^{s}n_i+1$ .
\\ 
\\
Eguch-Hori-Yang relation \cite{Eguchi:1995er}:
\bea{eghori}
 d^2\langle \sigma_{n}({\cal V}_{\alpha})\prod_{ i\in S}\sigma_{n_i}({\cal V}_{\alpha_i})\rangle_{0,d} = -2dn\langle \sigma_{n-1}({\cal V}_{\alpha+1})\prod_{ i\in S}\sigma_{n_i}({\cal V}_{\alpha_i})\rangle_{0,d}
\cr + \sum_{X\cup Y=S}\sum_{k=1}^{d}k^2n\langle \sigma_{n-1}({\cal V}_{\alpha})\prod_{ i\in X}\sigma_{n_i}({\cal V}_{\alpha_i}){\cal V}_{\alpha}\rangle_{0,d-k}\langle {\cal V}_{\alpha_j}\prod_{ j\in Y}\sigma_{n_j}({\cal V}_{\alpha_j})\rangle_{0,k}.
\eea
\\
Recursion relation among 2-point correlators:
\be{2pointr}
 \langle \sigma_{n}({\cal V}_{\alpha})\sigma_{m}({\cal V}_{\beta})\rangle = \frac{mnM_{\phi\sigma}\eta^{\delta\phi}\eta^{\gamma\sigma}}{(n+m+q_{\alpha}+q_{\beta})}\langle\sigma_{n-1}({\cal V}_{\alpha}){\cal V}_{\gamma}\rangle\langle\sigma_{n-1}({\cal V}_{\beta}){\cal V}_{\beta}\rangle
\ee
\\where
\bea{metric}
 \eta^{PQ} &=& \eta^{QP} =1,
\cr\eta^{PP} &=& \eta^{QQ} = 0,
\cr M_{PP} &= & M_{QQ} = 2.
\eea
\\
If we turn off all the couplings then we have $ \langle Q\rangle_{0,d} = 0, \langle P\rangle_{0,d} = 0 $ except 
$ \langle Q\rangle_{0,1} = 1 $. Then setting $n=2k_i, \alpha = 2$, and $ S = 0 $ in \eq{eghori} will give us the following 
equation

\be{1pointrel}
(k_i+1)^2\langle\sigma_{2k_i}(Q)\rangle_{0,k_i+1} = 2k_i(2k_i-1)\langle\sigma_{2k_i-2}(Q)\rangle_{0,k_i}
.\ee
\\
Therefore
\bea{1point}
\langle\sigma_{2k_i}(Q)\rangle_{0,k_i+1} &=&\frac{(2k_i)!}{(k_i+1)!(k_i+1)!}
\cr \langle\sigma_{2k_i-1}(Q)P\rangle_{0,k_i} &=& \frac{(2k_i-1)!}{(k_i!)^2}.
\eea
Also due to the selection rule \eq{gconl} we have for all $d$
\be{1point2}
\langle\sigma_{2k_i+1}(Q)\rangle_{0,d} = 0.
\ee
Plugging these into \eq{2pointr} will give the following 2-point correlator
\be{2pointco}
\langle\sigma_{2k_i-1}(Q)\sigma_{2k_j-1}(Q)\rangle_{0,k_i+k_j} = \frac{1}{4(k_i+k_j)}\frac{(2k_i)!}{(k_i!)^2}\frac{(2k_j)!}{(k_j!)^2}.
\ee 
To calculate the higher point correlators we can use \eq{trr} and reduce them to lower point correlators
by remembering the fact that the recursion relations are valid in the large phase space. For example 
\bea{example}
\langle\sigma_{2k_1-1}(Q)\sigma_{2k_2-1}(Q)\sigma_{2k_3}(Q)\rangle_{0,k_1+k_2+k_3} &=& 2k_3\langle\sigma_{2k_3-1}(Q)P\rangle_{0,k_3}\langle Q \sigma_{2k_1-1}(Q)\sigma_{2k_2-1}(Q)\rangle_{0,k_1+k_2}
\cr &=& \frac{1}{4}\frac{(2k_1)!}{(k_1!)^2}\frac{(2k_2)!}{(k_2!)^2}\frac{(2k_3)!}{((k_3!)^2}.
\eea
Repeated application of the recursion relations give us the following general correlators
\bea{topcorr1}
 \langle \sigma_{2k_1-1}(Q)\sigma_{2k_2-1}(Q)\prod_{i=3}^{n}\sigma_{2k_i}(Q)\rangle_{0,d} &= &\frac{d^{n-3}}{4}\prod_{i=1}^{n}\frac{(2k_i)!}{(k_i!)^2}
 \cr \langle \prod_{i=1}^{n}\sigma_{2k_i}(Q)\rangle_{0,d} &= & (d+1)^{n-3}\prod_{i=1}^{n}\frac{(2k_i)!}{(k_i!)^2}
 \cr  \langle P^2\prod_{i=1}^{n}\sigma_{2k_i}(Q)\rangle_{0,d} &=& d^{n-1}\prod_{i=1}^{n}\frac{(2k_i)!}{(k_i!)^2}
\eea
where $d=\sum_{i=1}^{n}k_i$. These results agree with the ones stated in \cite{norb}.

\bibliographystyle{JHEP}

\end{document}